\documentclass[a4paper,superscriptaddress,aip,apl,reprint]{revtex4-1}
\usepackage{graphicx}

\begin{document}

\title{High resolution cathodoluminescence hyperspectral imaging of surface features in InGaN/GaN multiple quantum well structures}
\date{\today}

\author{Jochen Bruckbauer}\email{jochen.bruckbauer@strath.ac.uk}\affiliation{Department of Physics, SUPA, University of Strathclyde, Glasgow G4 0NG, Scotland, United Kingdom}
\author{Paul R. Edwards}\affiliation{Department of Physics, SUPA, University of Strathclyde, Glasgow G4 0NG, Scotland, United Kingdom}
\author{Tao Wang}\affiliation{EPSRC National Centre for III-V Technologies, Department of Electronic and Electrical Engineering, University of Sheffield, Sheffield S1 3JD, United Kingdom}
\author{Robert W. Martin}\affiliation{Department of Physics, SUPA, University of Strathclyde, Glasgow G4 0NG, Scotland, United Kingdom}

\begin{abstract}
InGaN/GaN multiple quantum wells (MQWs) have been studied by using cathodoluminescence hyperspectral imaging with high spatial resolution. Variations in peak emission energies and intensities across trench-like features and V-pits on the surface of the MQWs are investigated. The MQW emission from the region inside trench-like features is red-shifted by approximately 45 meV and more intense than the surrounding planar regions of the sample, whereas emission from the V-pits is blue-shifted by about 20 meV and relatively weaker. By employing this technique to the studied nanostructures it is possible to investigate energy and intensity shifts on a 10 nm length scale.
\end{abstract}



\maketitle

The III-nitride alloy system, especially In$_x$Ga$_{1-x}$N, has become very attractive in the last two decades for opto-electronic applications, because of the tuneable band gap covering the entire visible and part of the ultraviolet spectrum. High brightness GaN-based light-emitting diodes \cite{Nak94} and laser diodes \cite{Nak96} already exhibit excellent optical properties and are commercially available, but there remains room for improvement due to a range of material challenges facing the growth of InGaN. Despite the introduction of advanced buffer layers the need to use lattice mismatched substrates means there is still a high threading dislocation (TD) density of up to $10^9$ cm$^{-2}$. A well-known defect in InGaN multiple quantum well (MQW) structures is the so-called V-pit or V-defect, where in general a TD terminates at the apex of an open hexagonal inverted pyramid with facets of $\left\{10\bar 11\right\}$-orientation \cite{Che98}. In addition, other nanoscale surface features can be present \cite{Tin03}, an example of which will be discussed in this letter. Dislocations can act as non-radiative recombination centres and thus can reduce the overall efficiency of the device, making it important to understand their luminescence properties. 

In this letter, we investigate samples exhibiting an additional type of macroscopic surface defect, which will be referred to as a trench-like feature, as well as the V-defect. Cathodoluminescence (CL) hyperspectral images with high spatial resolution were taken to study the luminescence behaviour of these features, including shifts in emission energy as a function of position on the surface of the MQW samples.

The investigated InGaN/GaN MQWs were grown on (0001) sapphire substrates by metal-organic chemical vapour deposition. To overcome the lattice mismatch a high-temperature atomically flat AlN buffer layer was employed \cite{Bai06}. After a 1 $\mu$m thick GaN layer 10 periods of InGaN/GaN layers were grown followed by a GaN capping layer of the same thickness as the GaN barriers. The surface morphology was investigated using a field emission gun scanning electron microscope (FEI Sirion 200), which was modified for CL spectroscopy \cite{Mar04}. The electron beam was scanned across the sample surface, while simultaneously acquiring a room temperature CL spectrum at each point on the surface using an electron multiplying charge-coupled device, resulting in a multi-dimensional data set, or hyperspectral image \cite{Chr91, Edw03}. The sample is tilted at 45$^{\circ}$ to the incident electron beam and the light is collected with a reflecting objective with its axis perpendicular to the beam. Typically an acceleration voltage of 5 kV, a beam current of $\approx$ 130 pA and a spot diameter of $\approx$ 2 nm were used. Afterwards the data set can be treated mathematically to extract images describing subsets of the recorded data. In the present case, each recorded spectrum was fitted by a Voigt function and the resulting fitting parameters (i.e. energy peak position, peak area, peak half-width) were plotted against the position on the sample surface to produce 2D maps. 

\begin{figure}[b]
\includegraphics[width=8.5cm]{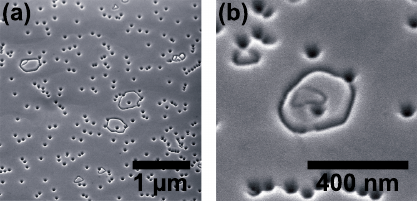}
\caption{\label{sem}(a) Typical SE micrograph of a 10 period InGaN/GaN MQW with trenches and V-defects; (b) High resolution SE micrograph showing a trench-like feature and the hexagonal character of the V-defect.}
\end{figure}

\begin{figure}[t]
\includegraphics[width=7.6cm]{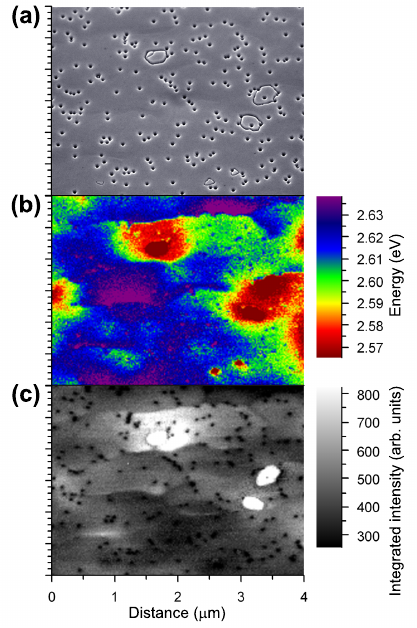}
\caption{\label{cl1}(Color online) CL hyperspectral information for a 10 period InGaN/GaN MQW: (a) SE image, (b) fitted CL energy map and (c) fitted CL integrated intensity map at RT.}
\end{figure}

The secondary electron (SE) image in Fig. \ref{sem}(a) reveals another type of macroscopic defect besides the well-known V-pit on the MQW surfaces. These features exhibit trench-like characteristics, which mostly form closed loops as seen in Fig. \ref{sem}(b). The trenches are in general thinner than V-pits and sometimes intersect them. The formation mechanism of the trenches is not clear and their optical characteristics are different. The main causes of the formation of the V-pit are not unambiguously known, but mostly likely are increased strain \cite{Kim98} and kinetic effects, i.e. a change of gallium incorporation and In segregation \cite{Wu98, Mah99}. The InN content is also reported to be an important factor; for low InN compositions there is almost always a TD originating at underlying interfaces or an inversion domain at the apex of the V-pit \cite{Wu98,Che98,Mah99}, whereas for high InN compositions ($>$30\%) stacking mismatch boundaries can also drive the formation \cite{Cho01}. Despite the appearance of the trench-like features and the presence of the V-pits with an average density in the lower 10$^9$ cm$^{-2}$ regime, the MQW samples still have excellent luminescence characteristics at RT \cite{Wan04}.

\begin{figure}[t]
\includegraphics[width=8.5cm]{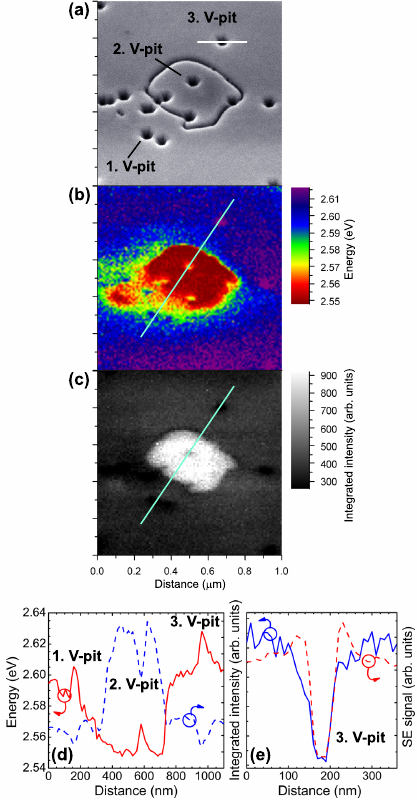}
\caption{\label{cl2}(Color online) High resolution CL hyperspectral images from sample in Fig. \ref{cl1}: (a) SE image, (b) fitted CL energy map, (c) fitted CL integrated intensity map at RT, (d) CL line scans through a trench-like feature with V-pits and (e) horizontal line scans through the third V-pit.}
\end{figure}

An advantage of CL hyperspectral imaging is the ability to map shifts in wavelength or energy, which can be traced back to features on the sample surface since an SE micrograph of the imaged area is recorded. Fig. \ref{cl1} shows a set of CL maps of a relatively large area containing three trench-like features and numerous V-pits. The luminescence is strongly influenced by the surface morphology. The peak energy inside the trench-like feature is red-shifted by about 45 meV and the emission is very intense in comparison to the surrounding planar regions. The V-pits show the opposite behaviour: the intensity is decreased and the emission energy is blue-shifted, which is more clearly shown in Fig. \ref{cl2}(b) and discussed in the following paragraph.

For a more detailed analysis, higher resolution hyperspectral images of the trench-like features were recorded, one of which is shown in Fig. \ref{cl2}. The SE micrograph in Fig. \ref{cl2}(a) displays a trench-like feature of roughly 500 nm diameter enclosing a few V-pits of about 65 nm in diameter. The CL energy and intensity maps for the fits are illustrated in Fig. \ref{cl2}(b) and (c), respectively. For another illustration of the luminescence behaviour of these features, providing information on a 10 nm length scale, line scans are taken across the trench-like feature and three V-pits (Fig. \ref{cl2}(d)). The line scans clearly demonstrate the opposite behaviour of the V-pits compared to the area inside the trench-like features. In particular, the peak energy for the V-pit in the middle of the trench-like feature is shifted by about 10 meV in energy over a distance of less than 25 nm. Fig. \ref{cl2}(e) shows horizontal line scans of the SE signal and CL integrated intensity of the third V-pit. The similar shape of the two scans is notable and the CL feature is broadened by no more than 10 nm compared to the SE scan.
 
The formation mechanism of the trenches is not clear. Due to their increased CL intensity, the trench-like features seem to localise the radiative carrier recombination while red-shifting the emission energy, which suggests an InN rich region or wider wells inside the trenches. Refs. \onlinecite{Flo03,Tin03} describe an increased and red-shifted CL emission coming from inclusions of increased height embedded in V-defects, which seem to be similar to the trench-like features in Fig. \ref{cl2}. They concluded that the bright emitting inclusions nucleate at the highly strained InGaN-to-GaN interface and attributed them to an increased InN content, where the V-defect induces InN segregation during the subsequent barrier and well growth. Similar features were also observed in Ref. \onlinecite{Kum07}, where it was also reported that a GaN barrier growth at higher temperatures than the InGaN well results in inclusion-free surfaces. A cause for the increased CL intensity inside the trench-like feature could be carrier localisation in quantum-dot-like structures of almost pure InN in the InGaN layer \cite{Odo99}, which would also explain the red-shifted emission as observed in Fig. \ref{cl2}. Furthermore, a difference in strain caused by the trench-like feature could assist InN migration and formation of these quantum dot (QD) structures within the trenches \cite{Lin00}.

The increase in emission energy from the V-pits can be caused by different effects, including decreased InN content, reduction in well width, increase in compressive strain and reduction of the quantum-confined Stark effect (QCSE). The above effects, however, are usually interrelated. For example, decreased InN content or narrower well widths reduce the influence of the QCSE, reinforcing the blue-shift. The increased energy emission from the V-pits is most likely due to a reduced InN incorporation into the InGaN MQWs along the sidewall of the pits, which is supported by results from InGaN single QW samples in Ref. \onlinecite{Yos09}. However, in Ref. \onlinecite{Wu98} it is also reported that the different crystal orientation of the sidewalls compared to the [0001]-direction leads to an increased InN corporation into the MQWs and an emission at lower energies was observed in contrast to the blue-shift in Fig. \ref{cl2}(b). It should be mentioned that the V-pits in both references are twice as large as the pits described here. It is difficult to determine which of the aforesaid mechanisms the dominant one is. However, it its clearly evident that V-pits act as non-radiative recombination centres since their luminescence intensity is greatly reduced.

In summary, CL hyperspectral imaging showed the influence of an additional kind of surface defect, the so-called trench-like feature, and the common V-pit on the luminescence on a 10 nm length scale. While V-pits shifted the emission energy towards higher energies, the emission from inside the trench-like features was red-shifted and more intense. This behaviour may be caused by different levels of InN incorporation or by the formation of QD-like structures leading to the localisation of carrier recombination inside the trenches.

The authors would like to thank UK EPSRC under Grant No. EP/H004157/1 and the University of Strathclyde for financial support.

\end{document}